\documentclass[prd,twocolumn,superscriptaddress,floatfix,amsmath,amssymb,amsfonts,longbibliography]{revtex4-2}

\usepackage{xcolor}
\usepackage{graphicx}
\usepackage{bm}
\usepackage{microtype}
\usepackage{physics} 
\usepackage[normalem]{ulem}
\usepackage{hyperref}
\usepackage{gensymb}
\usepackage{cleveref}

\definecolor{markus}{rgb}{0.09,0.65,0.63}

\definecolor{elinor}{rgb}{0.4, 0, 0.6}

\definecolor{jelte}{rgb}{0.3, 0.3, 0.9}

\definecolor{floriane}{rgb}{0, 0.722, 0.298}

\definecolor{hanyi}{rgb}{0, 0, 1}

\begin{document}

\title{Imaging Compressible and Incompressible Kinetic Energy of a 2d Bose-Einstein Condensate}

\author
{Elinor Kath$^\ast$$^{1}$, Jelte Duchêne$^{1}$, Hanyi Jang$^{1}$, Floriane Arrouas$^{1}$, Helmut Strobel$^{1}$ and Markus K. Oberthaler$^{1}$ \\ \quad \\
\normalsize{$^{1}$Kirchhoff-Institut f\"{u}r Physik, Universit\"{a}t Heidelberg\\
    Im Neuenheimer Feld 227, 69120 Heidelberg, Germany}\\ 
\normalsize{$^\ast$velocityfieldsforever@matterwave.de} \\ \quad \\}


\begin{abstract}
We introduce a method based on velocity-sensitive two-photon transitions to reconstruct the two-dimensional velocity field as well as the local atomic density of a Bose-Einstein condensate in a single realization. 
We quantitatively benchmark the method on a quantized vortex.
Applied to a condensate far from equilibrium, the method reveals the contrasting behaviors of the system's flow kinetic energy components, where the compressible component quickly decays while the incompressible one remains, dominating the long-time dynamics.
\end{abstract}
\maketitle
\noindent\textit{Introduction ---} 
In hydrodynamic descriptions, mass density $\rho(\mathbf r,t)$ and velocity $\mathbf v(\mathbf r,t)$ are fundamental fields: they specify the mass and momentum densities, $\rho$ and $\rho\mathbf v$, whose local balance laws constrain the fluid motion \cite{LandauLifshitz1987}.
Their joint spatial distribution determines mass transport and the flow kinetic energy density, $e_{\rm kin}=\rho|\mathbf v|^2/2$.
In the mean field description of weakly interacting atomic Bose-Einstein condensates (BECs), these fields are encoded in the amplitude and phase gradient of the order parameter $\Psi = \sqrt{n(\mathbf r)} e^{i\Phi(\mathbf r)}$, with $\rho=mn$ and $\mathbf v=(\hbar/m)\nabla\Phi$ \cite{Dalfovo1999}, where $\hbar$ is the reduced Planck's constant, $n$ is the particle density, and $m$ is the mass of the atoms.
\begin{figure}[b]
    \centering
     \includegraphics{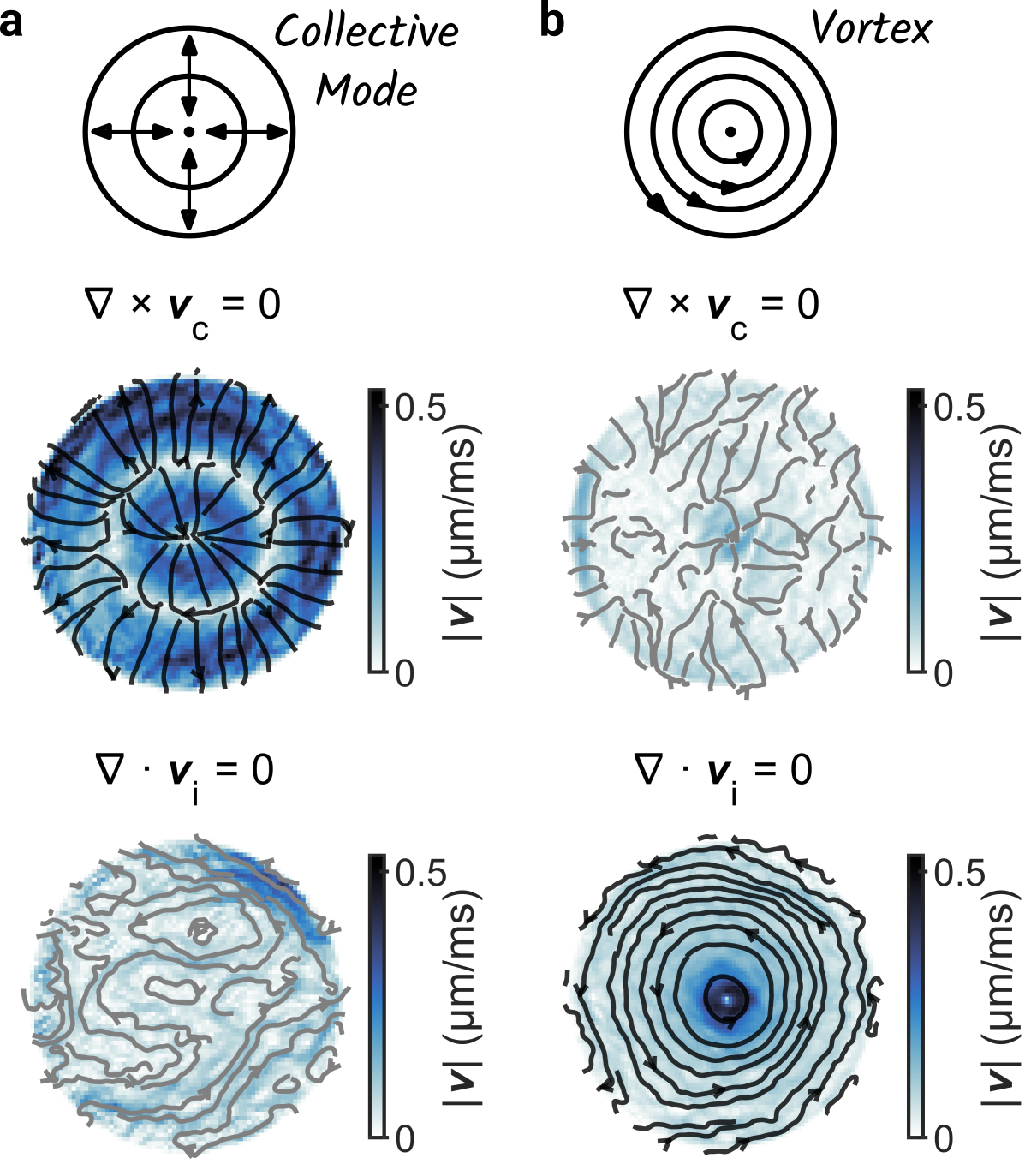} 
     \caption{\textbf{Helmholtz Components of Experimentally Obtained Velocity Fields of a Collective Mode and a Vortex.} The velocity fields of both examples are reconstructed from data averaged over 15 independent experimental realizations, and are then numerically decomposed into their Helmholtz components. \textbf{a)} Helmholtz decomposition of the reconstructed velocity field of a collective mode. The spatial structure of the excitation is mainly captured by the compressible component. \textbf{b)} In contrast, the velocity field of a vortex is predominantly incompressible.}
     \label{fig:Fig1}
\end{figure}

While density measurements are the standard way of analyzing dynamics of BECs, such as sound propagation \cite{Andrews1997,Christodoulou2021} and dispersive shock waves \cite{Hoefer2006}, complementary velocity-sensitive measurements have resolved coarse-grained rotational flow \cite{Muniz2006}, measured persistent currents via Doppler shifts of phonon modes \cite{Kumar2016}, identified the circulation signs of individual vortices \cite{Shin2017,Helmerson2019}, probed turbulent flow by tracking injected impurities \cite{Spielman2025}. 
Mass current extraction has been demonstrated in a discrete lattice \cite{Impertro2024LocalReadout}.
Nevertheless, only jointly measuring the \textit{in-situ} density and the full two-dimensional velocity field provides access to the spatially resolved kinetic energy density and its compressible and incompressible contributions. 
These observables offer a more complete characterization of turbulent flow, energy distribution across scales, and the dynamics following a quench \cite{Brachet1997KolmogorovinLowSuperflows,Gasenzer2011NonthermalFixedPoint,Bradley2026Turbulence}. 

In this Letter, we introduce an experimental method to extract both the atomic density and the velocity field.
Having access to the full velocity field, we can apply the Helmholtz decomposition to distinguish its compressible, i.e. rotation-free, and incompressible, i.e. divergence-free, components. 
This decomposition provides insight into the nature and spatial structure of the flow, for example by distinguishing sound-like excitations from vortical motion \cite{Lighthill1978}. 
Figure \ref{fig:Fig1} illustrates this distinction for two specific examples, showing the Helmholtz decomposition on experimentally obtained velocity fields of a collective mode (\ref{fig:Fig1}a) and a quantum vortex (\ref{fig:Fig1}b). 
While the collective mode has a predominantly compressible velocity field, the topological excitation in the form of a quantized vortex is dominated by the incompressible component.
\\\quad\\
\textit{Velocity-Sensitive Scattering --- } 
Our technique exploits the velocity dependence of two-photon scattering in a standing light wave, often referred to as Bragg scattering.
This principle has been used for a variety of applications in ultracold gases \cite{Bill1999CoherentSplittingBragg,Ketterle1999,Hagley2000MachZehnderBragg,Davidson2002ExcitationSpectrum,Bongs2003Spectroscopy,Sengstock2010SFinOpticalLatticesBragg,Zwierlein2021SpectroscopicProbesReview}.
We build on these ideas, probing the velocity along two axes by mapping local velocities to relative densities to reconstruct the full two-dimensional velocity field.

\begin{figure}[t]
     \centering
     \includegraphics{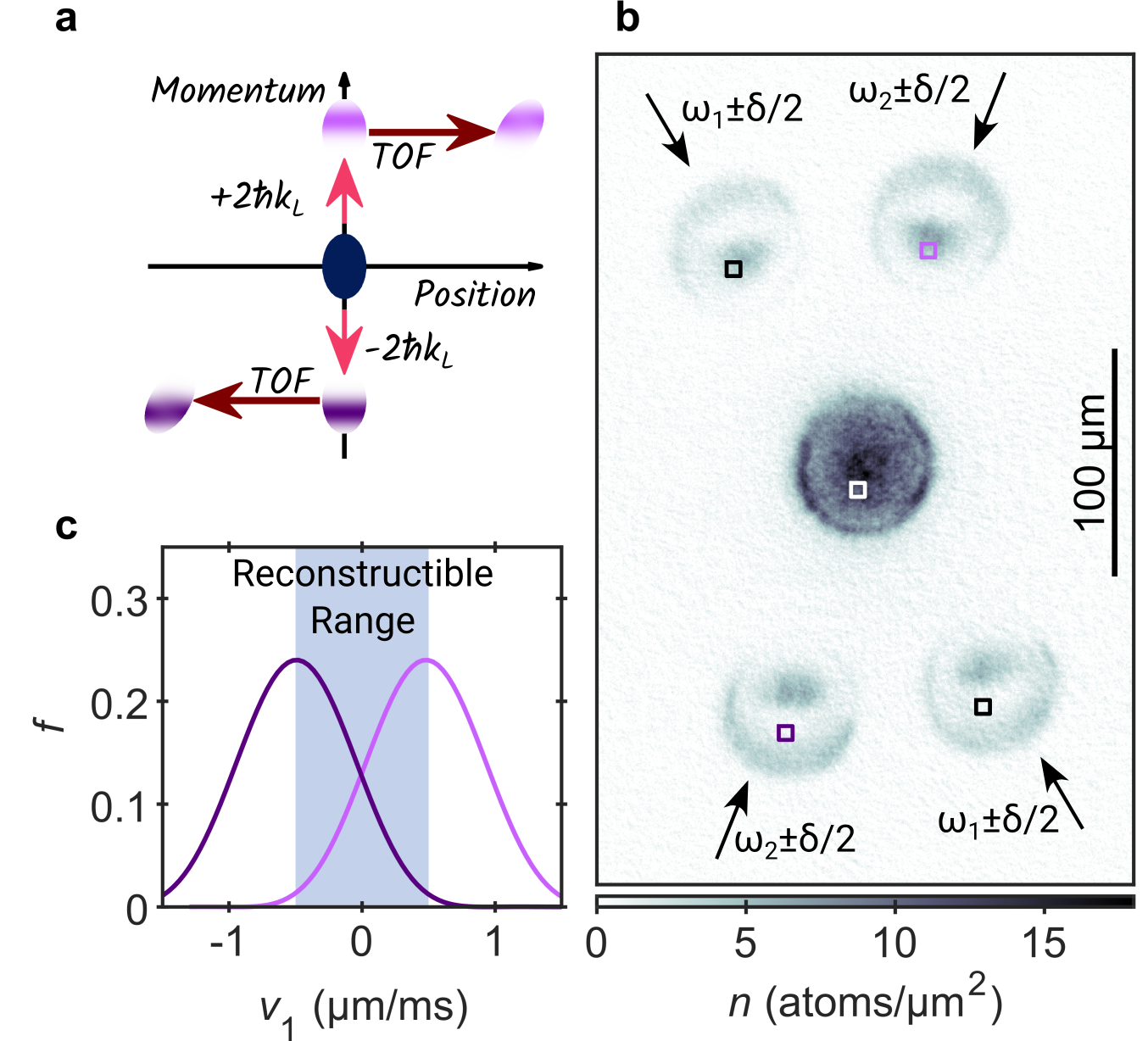} 
     \caption{\textbf{Momentum Transfer Through a Velocity-Sensitive Two-Photon Transition.} \textbf{a)} Schematic of the velocity-sensitive two-photon process and subsequent time-of-flight expansion. Atoms are transferred to momentum-shifted states depending on their velocity component along the interrogation axis. During TOF, the boosted clouds separate from the central cloud while their initial spatial distributions are sheared by their continued motion. \textbf{b)} Density distribution of a condensate with an excited box mode after simultaneous interrogation along two axes, showing the central condensate and four scattered-out clouds. \textbf{c)} Velocity dependence of the scattered-out fractions for the two-photon transitions along axis one at the chosen detuning. The shaded region indicates the range of velocity components along both interrogation axes that can be reliably reconstructed, $|v_{1,2}|<0.5\,\mu$m/ms.} 
     \label{fig:Fig2}
\end{figure}

As illustrated in a one-dimensional phase-space representation in \cref{fig:Fig2}a, the velocity-sensitive two-photon process addresses atoms depending on their velocity projections along the interrogation axes, transferring twice the photon recoil momentum $\pm2\hbar k_{\text{L}}$ while leaving the internal state of the atoms unchanged.
Afterwards, a short time-of-flight (TOF) spatially separates the boosted atoms from the initial cloud.
Established methods rely on a long TOF such that the shear of the phase-space distribution (see \cref{fig:Fig2}a) maps velocity to position, sacrificing spatial resolution.
Our technique retains most of the local information by reducing the necessary TOF duration and maps the velocities to local density ratios, i.e. out-coupled fractions.

We employ this principle in both directions along two axes by interfering counter-propagating laser light of two frequencies per beam, $\omega_{1,2}\pm\delta/2$, where $(1,2)$ denote the two interrogation axes.
Probing the velocity field with this configuration produces density distributions with four scattered-out clouds around the original BEC, as shown in \cref{fig:Fig2}b.

The light is pulsed on for an interrogation time $T_{\text{p}}$ with the temporal intensity envelope of a Hann pulse, $A(t)=\sin^2(\pi t /T_{\text{p}})$, to reduce sidebands \cite{Westbrook2025Hann}. 
For a sufficiently weak pulse, the probability for an atom to undergo scattering is proportional to the squared Fourier transform of the coupling (see Supplementary).
The out-coupled fraction is given by
\begin{equation}
    f(\Delta) =f_0\left[\frac{\sin(\Delta T_{\text{p}}/2)}
    {(\Delta T_{\text{p}}/2)\left[1-(\Delta T_{\text{p}}/2\pi)^2\right]}\right]^2.
    \label{eq:HannSquared}
\end{equation}
This spectral response is centered around $\Delta=0$, which sets the resonantly addressed velocity via
\begin{equation}
    \Delta = \frac{2\hbar |\bm{k}_{\text{L}}|^2}{m} + 2 \bm{k}_{\text{L}} \cdot \bm{v} - \delta.
\end{equation}
This dependence makes the encoding of the velocity information in the relative populations of the five resulting clouds possible.
We chose $\delta=2\pi\times 36\,$kHz, leading to a velocity dependent scattering probability as shown in \cref{fig:Fig2}c, defining our reconstructible velocity range of the components $v_1$ and $v_2$ along the interrogation axes, $|v_{\text{1,2}}|<0.5\,\mu$m/ms (shaded area).
For these velocities the reconstruction is robust because these velocities lead to a density signal in both scattered-out clouds.
\\\quad\\
\textit{Experimental Implementation --- }
We work with a quasi-two-dimensional BEC of approximately $6\times10^4$ $^{39}$K atoms at a scattering length around $100\,a_0$.
The dynamics is limited to the two-dimensional plane by a tight confinement along the gravity direction ($\omega_z = 2\pi\times 1.5\,$kHz).
The broad Feshbach resonance at 561\,G \cite{Hadzibabic2023} for the magnetic substate $F=1$, $m_F=-1$ allows to control the atom-atom interaction strength.
The in-plane confining dipole potential can be shaped and dynamically adjusted with blue-detuned laser light ($\lambda = 532\,$nm) using a Digital-Micromirror Device (DMD).

The setup consists of two interrogation axes aligned in the atomic plane, spanning a relative angle of $47\degree$.
The two frequencies on each axis are generated simultaneously through amplitude modulation via an acousto-optical modulator, and the light is retro-reflected after passing through the atoms.
The frequencies $\omega_{1,2}$ of the two axes are red-detuned by $500\,$GHz from the atomic $D1$ line at $770\,$nm to suppress single-photon processes, and are detuned by $30\,$MHz relative to each other in order to prevent interference.

To reduce density distortions from atom-atom interactions, the scattering length is reduced to zero in approximately 150$\,\mu$s before the interrogation pulses. 
Simultaneously, the in-plane confinement of the DMD is ramped down.
Then, the pulses are applied simultaneously for a duration of $T_{\text{p}}=540\,\mu$s.
After a TOF of $4.5\,$ms the atoms are imaged using absorption imaging \cite{Maurus2021}.
\\\quad\\
\textit{Reconstructing 2d Velocity Fields --- }
To extract the spectral response, we scan the detuning $\delta$ and record the scattered-out fraction $f$ in each cloud on a BEC at rest.
From this we obtain the detuning resonant to $v_{1,2}=0$, $\delta = 2\pi\times34.6\,$kHz, the response amplitude related to the coupling strength of the states, and the spectral width related to the duration of the interrogation pulse (see Supplementary).
For the following measurements we chose a fixed detuning of $\delta = 2\pi\times36\,$kHz.

Since the densities in the boosted clouds are not independent, the velocities $v_{1,2}$ cannot be reconstructed separately.
To consider the inter-dependencies of the scattered-out clouds via depletion of the center cloud, the parameters extracted from the spectral response, together with the chosen detuning, define the parameters of the Hamiltonian that models the four two-photon transitions (see Supplementary). 
With this quantitative model we calculate a lookup map for each of the four boosted clouds which contains the expected scattered-out fraction $f$ for all pairs of velocities $v_1$ and $v_2$.
Finally, a velocity field is extracted from data as in \cref{fig:Fig2}b: 
For each position in the cloud, the scattered-out fractions are compared to the lookup maps, and the most likely set of velocities $(v_1,v_2)$ is assigned to the pixel.
For this, we improve the signal-to-noise ratio by binning the density images in bins of $2\times2$, leading to pixels covering $(0.91\,\mu\text{m})^2$ in the atom plane.

The readout can also be viewed as a generalized quantum measurement. 
The initially unoccupied momentum states provide auxiliary modes to which the low-momentum atomic field is coherently coupled by the measurement pulses. 
Spatially resolved population detection realizes a projective measurement on the enlarged mode space and hence a positive operator-valued measure (POVM) on the initial field. 
This construction is analogous to the auxiliary-state spin readout \cite{Kunkel2019} and related phase-space protocols \cite{Weitenberg2026PhaseSpaceInLattices}. 
\\\quad\\
\textit{The Benchmark: A Quantum Vortex --- } 
\begin{figure}
     \centering
     \includegraphics{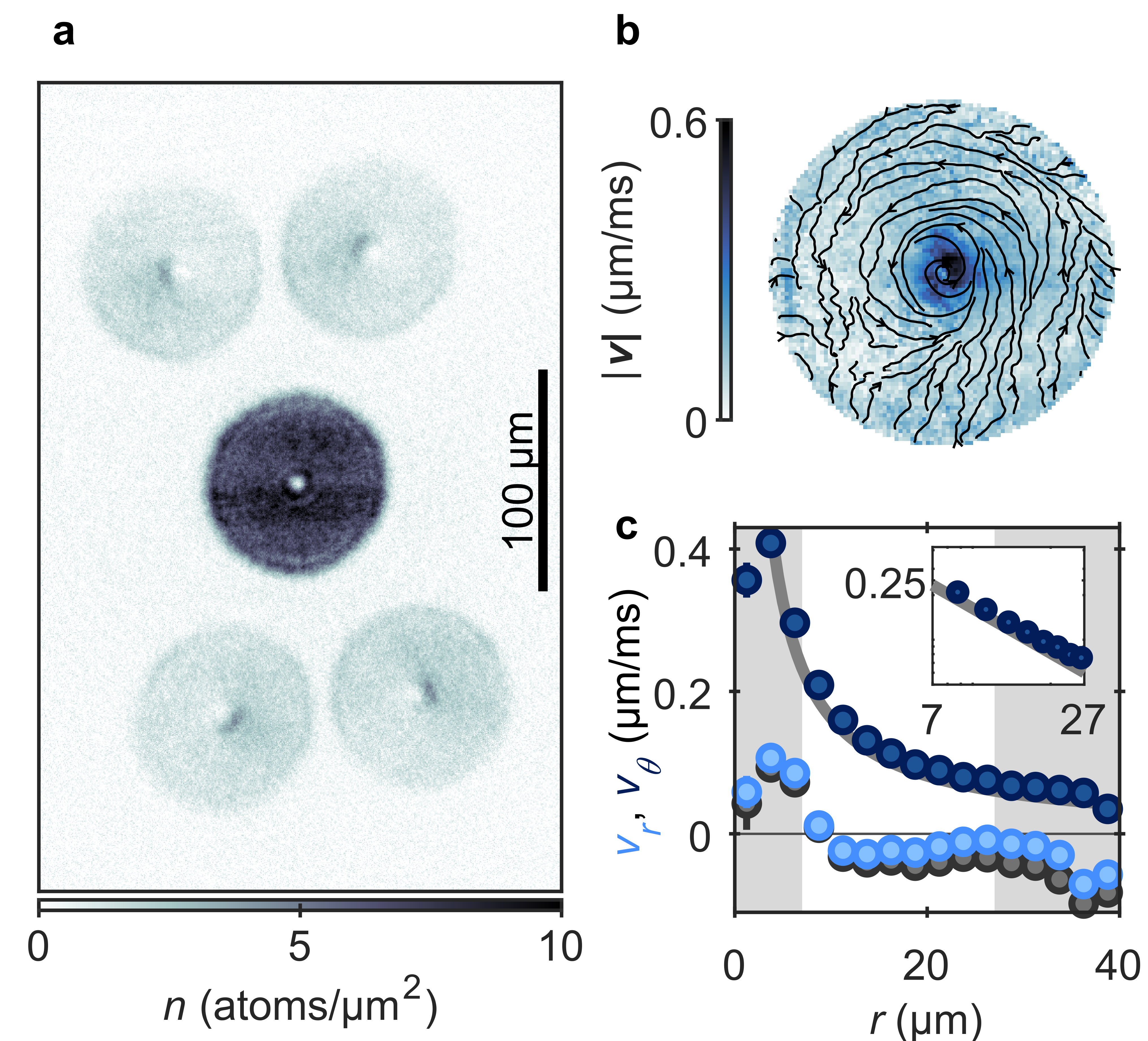} 
     \caption{\textbf{Benchmarking the Technique on a Single Quantum Vortex.} 
     \textbf{a)} Density distributions to reconstruct the velocity field of a BEC with a quantum vortex in the center. The absorption image was averaged over 15 independent experimental realizations after post-selecting on the vortex position. The central density depletion identifies the vortex core. \textbf{b)} Reconstructed in-plane velocity field. The background color shows the velocity magnitude, while the overlaid streamlines trace the local flow direction. The streamlines are numerically generated from the reconstructed velocity components and serve to visualize the circulating flow around the vortex. The contribution from the residual harmonic confinement is subtracted. \textbf{c)} Radial and azimuthal velocity components, $v_r$ (light blue, gray points) and $v_\theta$ (dark blue points), as a function of distance from the vortex center. The azimuthal component exhibits the expected $1/r$ dependence. The gray points show the radial velocity before subtracting the velocity contribution of the harmonic trap, while the light-blue points show the residual radial velocity after subtraction. The remaining radial component arises from dynamics induced during the measurement process.}
     \label{fig:Fig3}
\end{figure}
To benchmark the technique we create a single vortex in the center of the condensate using the Chopstick method \cite{Anderson2016,Kwon2021VortexCollider}, wait for $1\,$s to allow any inadvertent excitations to damp out, and perform a velocity measurement.
We minimize the influence of imaging noise for this demonstration by averaging the absorption images of 15 experimental realizations, post-selected on the vortex position.
The position and sign of the vortex is identified by calculating the curl of the individual velocity fields and identifying local maxima and minima (see Supplementary).
The resulting density distributions are shown in \cref{fig:Fig3}a. 
The core of the vortex is directly visible as a density depletion in the center of each cloud.
Reconstructing the velocities yields the field shown in \cref{fig:Fig3}b, with the velocity contribution induced by a shallow harmonic trapping potential subtracted (see below).
Decomposing the field into radial and angular velocity components, $v_{r}$ and $v_{\theta}$, shown in \cref{fig:Fig3}c, reveals the expected $1/r$ dependence of the angular velocity associated with a quantum vortex.

The reconstructed velocity components $v_{\theta}$ and $v_{r}$ agree with the analytical prediction except in the regions close to the vortex core and at the edge of the condensate. 
The observed deviations originate from the dynamics induced by the reduction in interaction strength before the measurement pulses. 
This changes the balance of forces in the cloud, leading to a contraction in the residual harmonic confinement and to additional quantum-pressure-driven flows. 

First, a shallow, residual harmonic potential stemming from the curvature of the magnetic field ($\omega_{\text{harmonic}} \approx 2\pi \times 5\,$Hz) causes the cloud to slowly contract. 
This produces a large-scale inward radial velocity. 
Since this contribution is deterministic, it can be independently characterized and subtracted from the measured velocity field, as done for the data shown in \cref{fig:Fig3}.
The gray points in \cref{fig:Fig3}c show the radial component if the harmonic contribution is not corrected for.

Second, the interaction quench changes the chemical potential and hence the system's balance between interaction and quantum-pressure contributions. 
The resulting density redistribution generates small-scale flows, visible in \cref{fig:Fig3}a as density ripples propagating outwards from the vortex core and inwards from the condensate edges. 
Consequently, the core appears substantially larger than expected from the \textit{in-situ} healing length of $\xi=0.6\,\mu$m.
The quantum-pressure contribution leads to a velocity change due to the acceleration $a_{\mathrm{QP}} =\hbar^2\,\nabla (\nabla^2\sqrt{n}/\sqrt{n})/2m^2$.
The velocity induced by a density variation with a characteristic length scale $L$ is proportional to $v_{\mathrm{QP}} \propto 1/L^3$
and thus density variations on short length scales, such as a vortex core or the condensate edge, can produce appreciable local accelerations (see Supplementary). 
For further quantitative analysis we exclude the gray shaded regions in \cref{fig:Fig3}c, where the quantum pressure dynamics lead to strong density modulation.
\\
\quad\\
\textit{Compressible \& Incompressible Kinetic Energies} --
By reconstructing both the spatial density and velocity distributions, we can estimate the local kinetic energy density,
\begin{equation}
    e=\frac{\hbar^2|\nabla\Psi|^2}{2m} =
	\underbrace{\frac{\hbar^2}{2m}|\nabla\sqrt n|^2}_{e_q}
	+\underbrace{\frac{m}{2}|\mathbf u_c|^2}_{e_c}
	+\underbrace{\frac{m}{2}|\mathbf u_i|^2}_{e_i}.
\end{equation}
We can separate its compressible, $e_c$, and incompressible,  $e_i$, contributions by decomposing the density-weighted velocity field $\mathbf{u}(\mathbf r) = \sqrt{n(\mathbf r)} \mathbf v(\mathbf r)$ into $\nabla\times\mathbf u_c=0$ and $\nabla\cdot\mathbf u_i=0$ with the Helmholtz decomposition \cite{TsubotaLvov2010EnergyComponents,BradleyAnderson2012EnerySpectra}.
\begin{figure}
     \centering
     \includegraphics{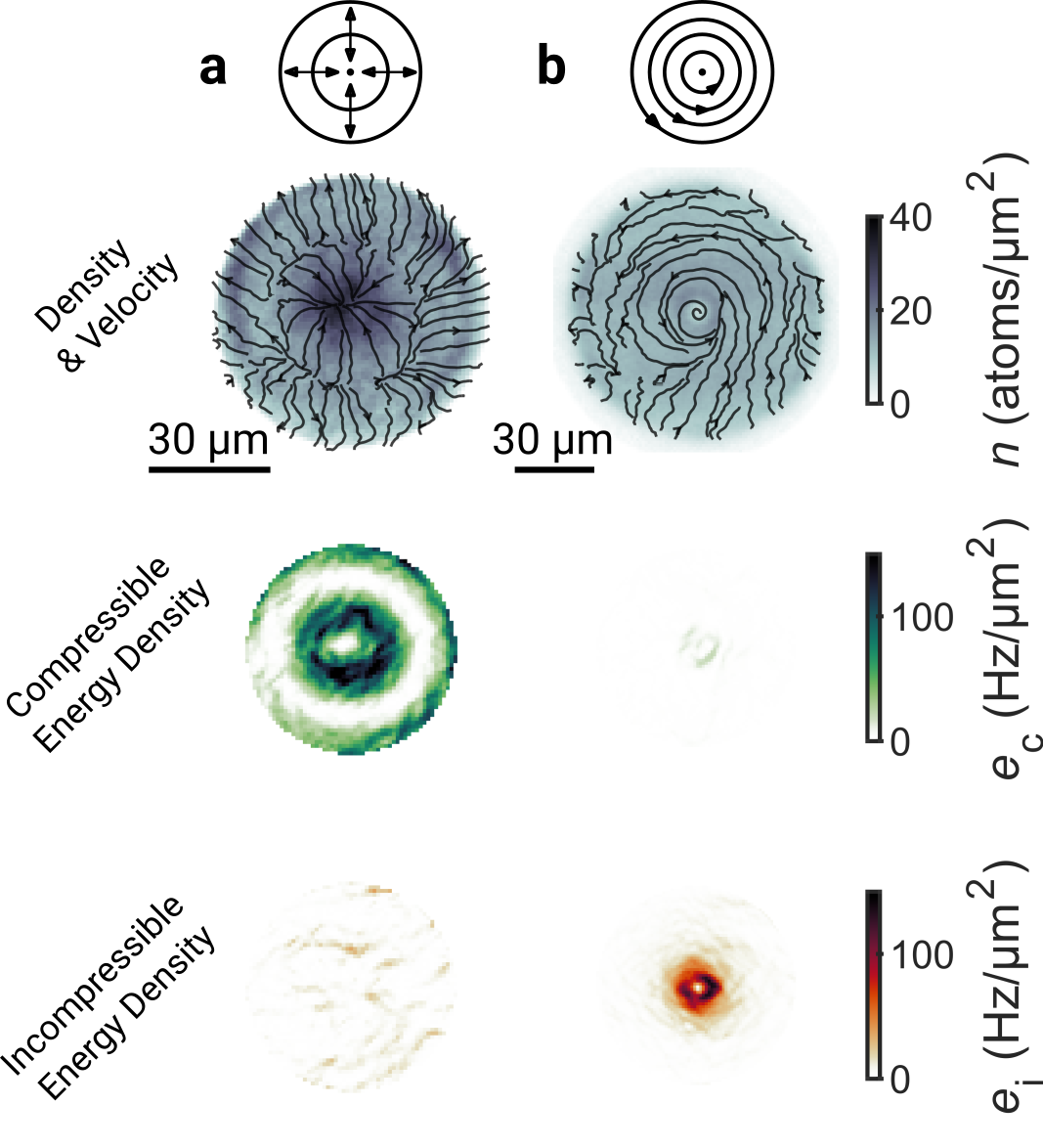} 
     \caption{\textbf{Spatially Resolved Compressible and Incompressible Kinetic Energy of a Collective Mode and a Vortex.} The top row shows the estimated density distributions, overlaid with streamlines of the reconstructed velocity fields. The density distributions of 15 independent experimental realization have been averaged for both examples before the velocity fields and densities were reconstructed. The middle and bottom rows show the corresponding compressible and incompressible contributions to the flow kinetic energy density, respectively. \textbf{a)} Collective mode. The kinetic energy is predominantly compressible, with its spatial distribution reflecting the structure of the excitation. \textbf{b)} Quantized vortex. The flow is dominated by incompressible kinetic energy, concentrated around the vortex core where the velocity is high and the density remains finite. The small compressible contribution is localized near the core and is attributed to an artifact arising from quantum-pressure effects in the measurement process.}
     \label{fig:Fig4}
\end{figure}

To estimate the \textit{in situ} atomic density distribution, we re-combine the four scattered-out clouds with the central cloud by adding their respective density distributions (\cref{fig:Fig4}, top row). 
This reconstruction builds on that atomic motion during the $4.5\,$ms TOF is negligible, such that the measured distributions retain their initial spatial structure. 
We calculate the compressible and incompressible kinetic energy densities of the collective mode and the quantum vortex and visualize their spatial distributions in \cref{fig:Fig4}, second and third row.
As expected, the collective mode (\cref{fig:Fig4}a) carries predominantly compressible kinetic energy, whereas the vortex flow (\cref{fig:Fig4}b) is dominated by incompressible kinetic energy. 
\begin{figure*}[!t]
     \centering
     \includegraphics{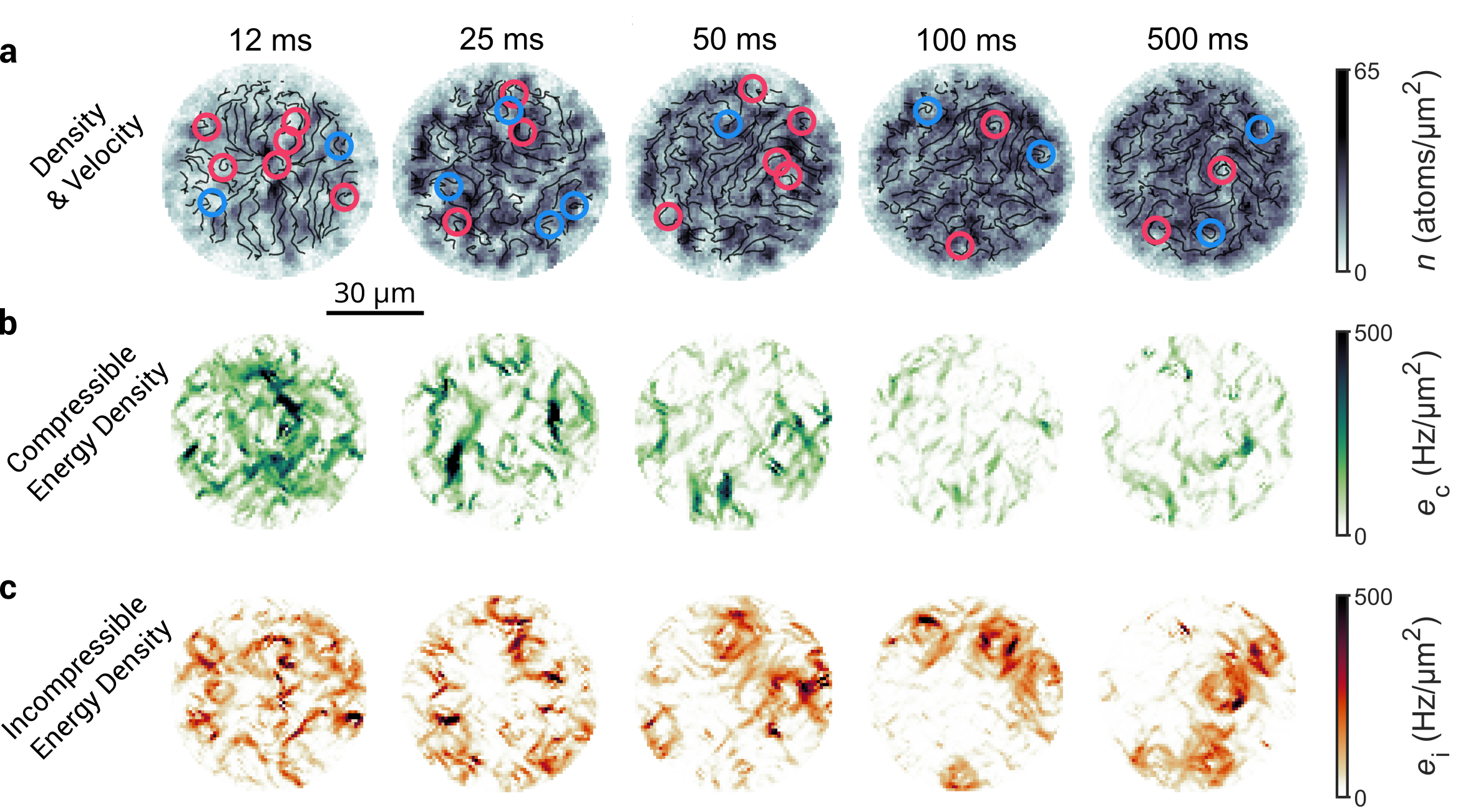} 
     \caption{\textbf{Spatially Resolved Flow Kinetic Energy Components in a Far-From-Equilibrium Condensate.} We drive the system out of equilibrium by imprinting a spatially structured phase and follow the subsequent dynamics using single-shot measurements. \textbf{a)} Reconstructed density distributions overlaid with streamlines of the velocity field; vortex positions and circulation signs are marked by circles (red: positive charge, blue: negative charge). \textbf{b)} Compressible kinetic energy density, \textbf{c)} Incompressible kinetic energy density. At early times, the density exhibits strong spatial modulations, accompanied by localized, propagating sound-like structures in the compressible component. Vortices appear as localized features in the incompressible component. At later times, density fluctuations become less pronounced and vortices begin to annihilate. The compressible kinetic energy becomes more spatially distributed and decays, while the incompressible contribution remains concentrated around the vortices.}
     \label{fig:Fig5}
\end{figure*}

For a quantitative comparison with theoretical expectations, we integrate the experimentally inferred kinetic energy densities, omitting the boundary areas influenced by the quantum pressure.
In the following, energies are expressed in frequency units (Hz) and can be converted to Joules by multiplying by Planck's constant $h$.
For the collective mode at this specific time we find a compressible kinetic energy of $E^{\text{coll.}}_{\text{c}}\approx 80\,$kHz, while for the system with the vortex it is as small as $E^{\text{vortex}}_c\approx 4.5\,$kHz.

It is interesting to note that the energy of a vortex is the sum of incompressible energy, interaction energy, and energy associated to quantum pressure \cite{Bradley2012EnergySpectra}.
The latter is transformed into curl-free velocities when switching off the interactions, and with that becomes accessible in our measurements.
The quantum pressure energy is predicted to be $E_Q = 3.4\,$kHz, numerically obtained from a static vortex profile.
We observe this energy (\cref{fig:Fig4}b, second row) concentrated around the vortex core. 

The prediction for the incompressible kinetic energy for the given vortex is \cite{PethickSmith2008}
\begin{align*}
E^{\text{analytic}}_i
    &= \frac{h n }{4\pi m}
\ln\left(\frac{27\,\mu\mathrm{m}}{7\,\mu\mathrm{m}}\right)\\
    &= (16\pm3)\,\mathrm{kHz},
\end{align*}
where the integration area is given by the unshaded area in \cref{fig:Fig3}c, and the error stems from the uncertainty in the estimated atomic density via speed of sound measurements. 
Integrating the experimentally obtained incompressible kinetic energy density yields $E^{\text{vortex}}_{i} = 21\,\mathrm{kHz}$.
In our observation, this energy density is not rotationally symmetric as expected. 
In order to estimate our systematic deviations we analyzed the incompressible kinetic energy in twelve angular sectors, finding a maximal deviation of $27\%$.
Since the kinetic energy depends on both the local density and velocity, the incompressible kinetic energy of the vortex is largest around the core and approaches zero inside it. 
\\\quad\\
\textit{Velocity Fields from Single Realizations} ---
Finally, we apply our technique to a condensate far from equilibrium. 
Because these dynamics are not deterministic, averaging density distributions across independent experimental realizations would obscure the evolution. 
Instead, we reconstruct the velocity field from each single-shot measurement. 
Since the density ratios are sensitive to imaging noise, this can lead to isolated pixels with unphysical results of the velocity reconstruction. 
To obtain faithful single-shot velocity fields, we identify outliers using a local median absolute deviation criterion and replace them by biharmonic interpolation (see Supplementary).

For state preparation we flash on a spatially structured optical potential for $500\,\mu$s, generated by the same DMD that shapes the box trap.
The pattern is obtained from uncorrelated single-pixel noise drawn from a uniform distribution (the same for all realizations, i.e. quenched disorder), which is low-pass filtered in Fourier space with a sharp cutoff that corresponds to a wavelength of $20\,\mu$m in the atomic plane.
The resulting phase imprint predominantly injects compressible excitations, similar to the random-phase initial states studied numerically in \cite{TsubotaLvov2010EnergyComponents}.
If enough energy has been injected, vortices emerge on a time scale of $\tau\propto1 / \mu < 1\,$ms, where $\mu$ is the chemical potential.
Representative single-shot velocity fields and corresponding kinetic energy distributions are shown in \cref{fig:Fig5} for different evolution times. 

The strongly perturbed density distribution prevents unambiguous identification of vortex cores from the density alone.
The position and sign of vortices can be identified reliably from the velocity fields (see Supplementary) and are marked in \cref{fig:Fig5}a. 
At early times, the atomic density exhibits strong spatial modulations, accompanied by compressible excitations visible as localized, propagating sound-like structures (\cref{fig:Fig5}b). 
Incompressible excitations (\cref{fig:Fig5}c) appear as localized features. 
At later times, the density fluctuations become less pronounced, and vortices begin to annihilate.
The compressible kinetic energy becomes more spatially distributed and decays, while the incompressible component remains localized around the vortices, and thus is decisive for long-time coarsening dynamics \cite{Rasch2026}.
\\\quad\\
\textit{Conclusion -- }
We have demonstrated a method based on velocity-sensitive two-photon transitions that enables the reconstruction of two-dimensional velocity fields together with the corresponding density distributions of a quasi-two-dimensional Bose-Einstein condensate. 
By probing the velocity along two axes, our approach retains spatial information while providing access to the full in-plane flow field from individual experimental realizations. 
Combining the reconstructed density and velocity fields and applying a Helmholtz decomposition allows us to spatially resolve the compressible and incompressible contributions to the flow kinetic energy density in position space. 
Applying the method to a condensate driven far from equilibrium, we access the flow kinetic energy on a single-shot basis.
This provides a route to quantitatively study the spatial redistribution of kinetic energy and the coupled dynamics of sound and vortices in quantum fluids \cite{TsubotaLvov2010EnergyComponents, Anderson2013ExpTurbulence,Zoran2016TurbulentCascade,Bradley2025VelocityCorrelations,Bradley2025SteadyStateTurbulence}.

\section*{Acknowledgements} 
We thank Niklas Rasch for sharing his theoretical expertise in far-from-equilibrium dynamics, Andreea Oros for making the numerical simulations possible, and Louis Evans and Nikolas Liebster for their contributions to the experimental activities at the beginning of the project.
This work is supported by the Deutsche Forschungsgemeinschaft (DFG, German Research Foundation) under Germany’s Excellence Strategy EXC 2181/1 - 390900948 (the Heidelberg STRUCTURES Excellence Cluster) and under SFB 1225 ISOQUANT - 273811115.

\bibliography{Bibliography}

\clearpage
\section*{Supplementary}

\subsection*{Reconstructing Velocities from Density Ratios}
\begin{figure*}
    \centering
    \includegraphics[width=1\linewidth]{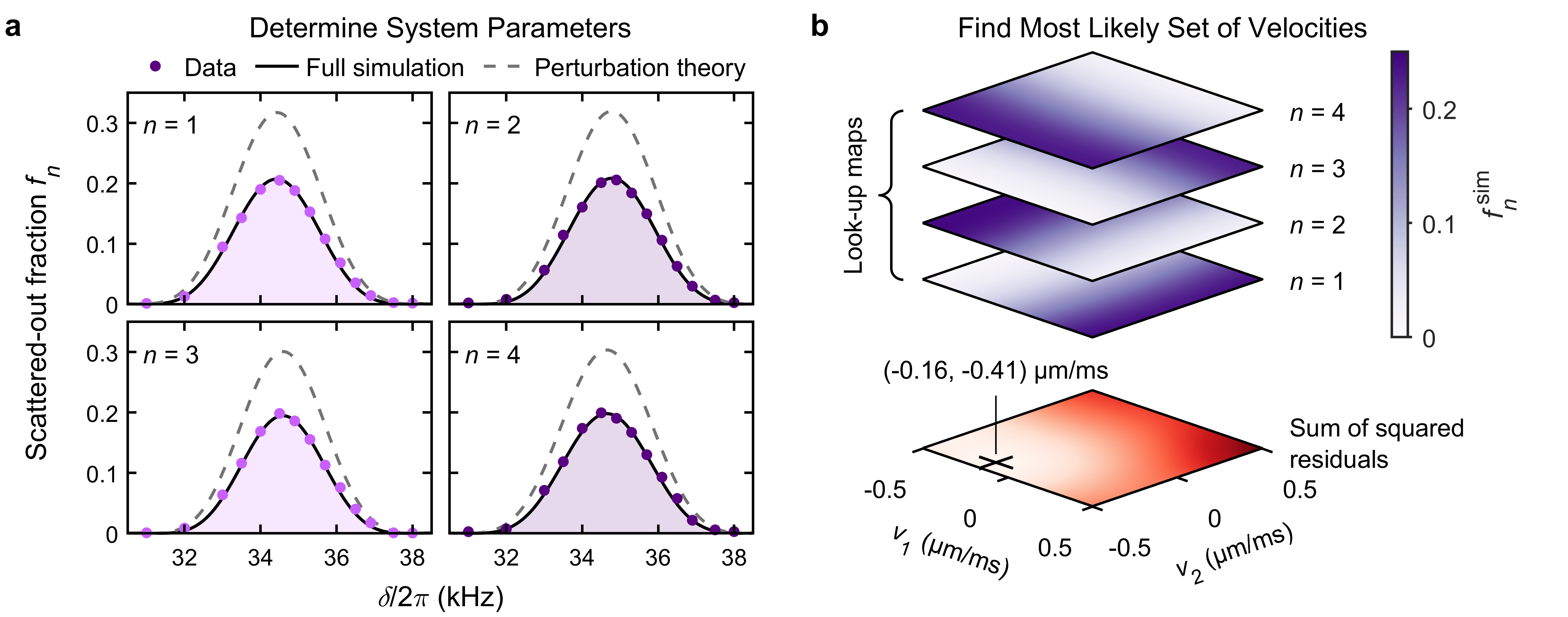}
    \caption{\textbf{Calibration and velocity extraction.}
    \textbf{a)} Measured scattered-out fractions $f_n$ of the four clouds of a condensate at rest as a function of the detuning $\delta$.
    Clouds $n=1,2$ ($n=3,4$) belong to interrogation axis 1 (2).
    Solid lines show the full model for the calibrated parameters, dashed lines first-order perturbation theory.
    Perturbation theory overestimates the scattered-out fractions since it neglects the competition of the four directions for the center condensate.
    \textbf{b)} Lookup maps $f_n^{\text{sim}}(v_1,v_2)$ for the detuning $\delta/2\pi = 36\,$kHz used for the velocity measurements, and sum of squared residuals for the example measured fractions $f_n^{\text{exp}} = (0.11, 0.2, 0.11, 0.23)$ of a single pixel.
    Its minimum (marked by a cross) yields $(v_1,v_2) = (-0.16, -0.41)\,\mu$m/ms.}
    \label{fig:SuppLookUp}
\end{figure*}
To generate the lookup maps used for velocity reconstruction, we model the scattering process as four velocity-sensitive two-state transitions: one for each propagation direction along each of the two measurement axes. 
All four transitions share the same initial state, corresponding to the unscattered condensate, but couple it to four different momentum states. 
This is captured by the Hamiltonian
\begin{equation}
    \hat{H}(t) = \sum_{n=1}^{4} A(t)\frac{ \hbar\,\Omega_n}{2} \left( e^{-i \Delta t} |\boldsymbol{v} \rangle \langle \boldsymbol{v} + 2\hbar k_L\boldsymbol{e}_n/m|
        + \text{h.c.}
    \right),
    \label{eq:Hamiltonian}
\end{equation}
where $\boldsymbol{e}_n$ denotes the outcoupling directions.
The Rabi frequencies of the four two-state transitions, $\Omega_n$ and the resonance for stationary atoms are extracted from fits to the system's spectral response.

We measure the spectral response on a stationary condensate by scanning the detuning $\delta$ and recording the scattered-out fractions $f_n$ in each of the four boosted clouds, see \cref{fig:SuppLookUp}a.
We fit the resulting curves using the function given in \cref{eq:HannSquared}, which is motivated by perturbation theory:
For one two-state transition and weak coupling, $\Omega_n T_{\text{p}} \ll 1$, the scattered-out fraction is given by the squared modulus of the Fourier transform of the pulse envelope.
However, the four two-state transitions compete for the same atoms in the BEC. 
This leads \cref{eq:HannSquared} to overestimate the scattered-out fractions (see dashed lines in \cref{fig:SuppLookUp}a).
We therefore use the spectral response only to extract the system parameters, but solve the model describing the full system, \cref{eq:Hamiltonian}, to compute the lookup maps.
Each lookup map contains the expected scattered-out fractions for pairs of velocity components, calculated as 
\begin{equation}
    f_n^{\mathrm{sim}}(v_1,v_2)= 
    \left|
    \left\langle
    \boldsymbol{v} + 2\hbar k_L\boldsymbol{e}_n/m
    \left|
   e^{-\frac{i}{\hbar}\int_0^{ T_{\text{p}}}\hat{H}(t)\,\mathrm{d}t}
    \right|
    \boldsymbol{v}
    \right\rangle
    \right|^2,
\end{equation}
as can be seen in \cref{fig:SuppLookUp}b.
The size of the look-up maps sets the largest extractable velocity in each axis, set to $0.5\,\mu$m/ms for the data shown in this letter.

Reconstructing the velocity field from the density distributions of a velocity-sensitive measurement is done pixel-bx-pixel.
Pixels corresponding to the same location in all five clouds are identified, and the local scattered-out fractions $f_n^{\mathrm{exp}}$ in each cloud are computed. 
The location gets assigned the velocity components $(v_1,v_2)$ which best agree with scattered-out fractions, see \cref{fig:SuppLookUp}b. 
This is determined by minimizing the sum of squared residuals with respect to the lookup maps,
\begin{equation}
(v_1,v_2) = \underset{v_1,v_2}{\operatorname{argmin}}  \sum_{n=1}^{4} \left( f_n^{\mathrm{exp}} - f_n^{\mathrm{sim}}(v_1,v_2) \right)^2.
\end{equation}
\\\quad\\
\subsection*{Identifying Vortex Signs}
\begin{figure}
     \centering
     \includegraphics{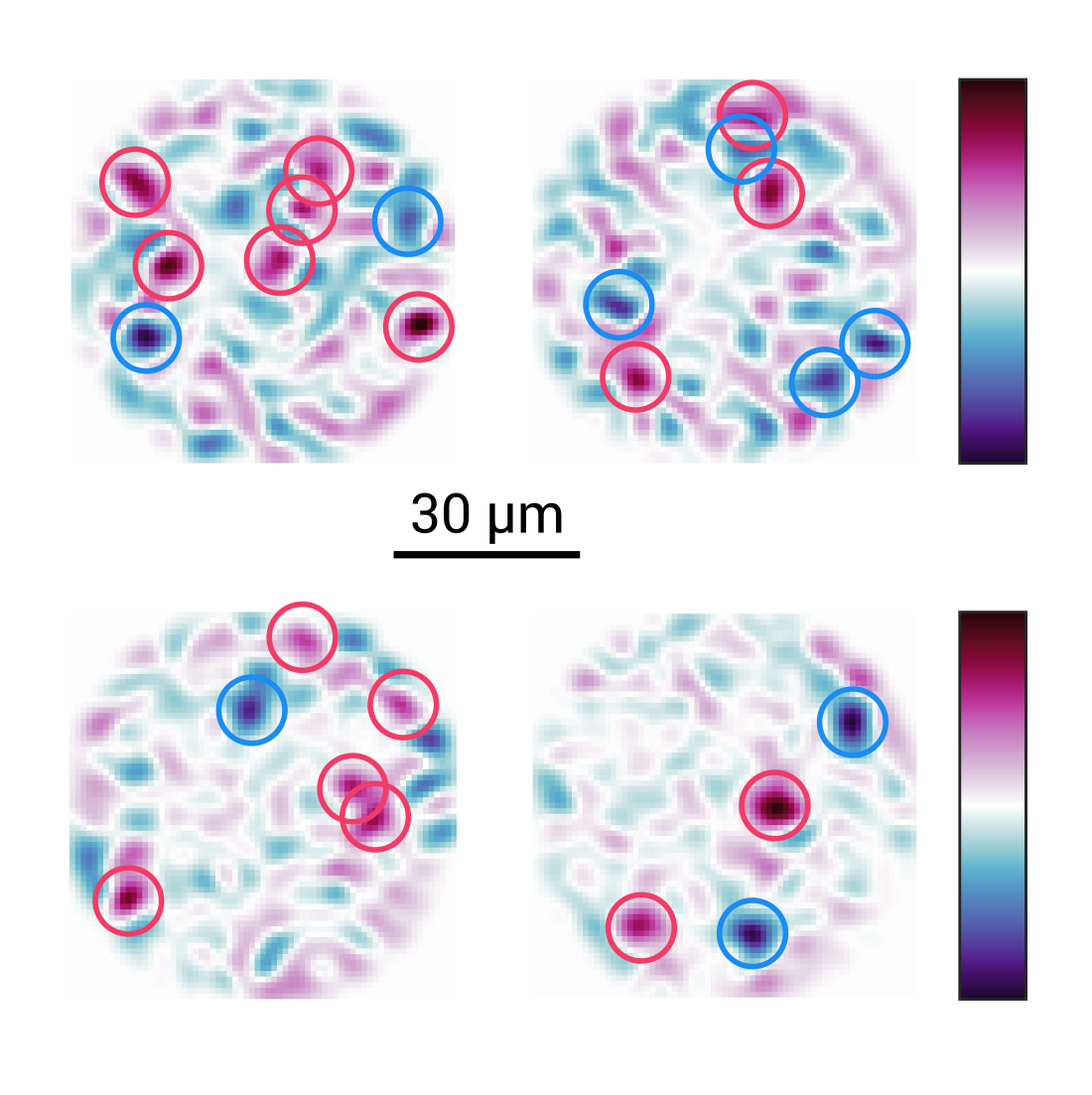} 
     \caption{\textbf{Identifying Vortices in the Curl of a Velocity Field.} Shown is the curl of the denoised velocity field after applying a Gaussian filter with a standard deviation of 2 pixels. Vortices appear as local maxima and minima in the curl, corresponding to their respective charges. They are identified by applying a threshold to the absolute value of the curl.}
     \label{fig:FindVorzis}
\end{figure}
To locate vortices and determine their charge, we examine the curl of the denoised velocity fields after applying a Gaussian filter with a standard deviation of 2 pixels. Vortices appear as local maxima and minima in the curl (see \cref{fig:FindVorzis}), corresponding to positive and negative circulation, respectively. We identify them by applying a cutoff threshold and selecting extrema whose absolute values exceed this threshold. The experimental realizations shown in \cref{fig:FindVorzis} correspond to those shown in \cref{fig:Fig5} for evolution times $12,25,50,500\,$ms.
\\\quad\\
\subsection*{GPE Simulations}
\begin{figure}
     \centering
     \includegraphics[width=1\linewidth]{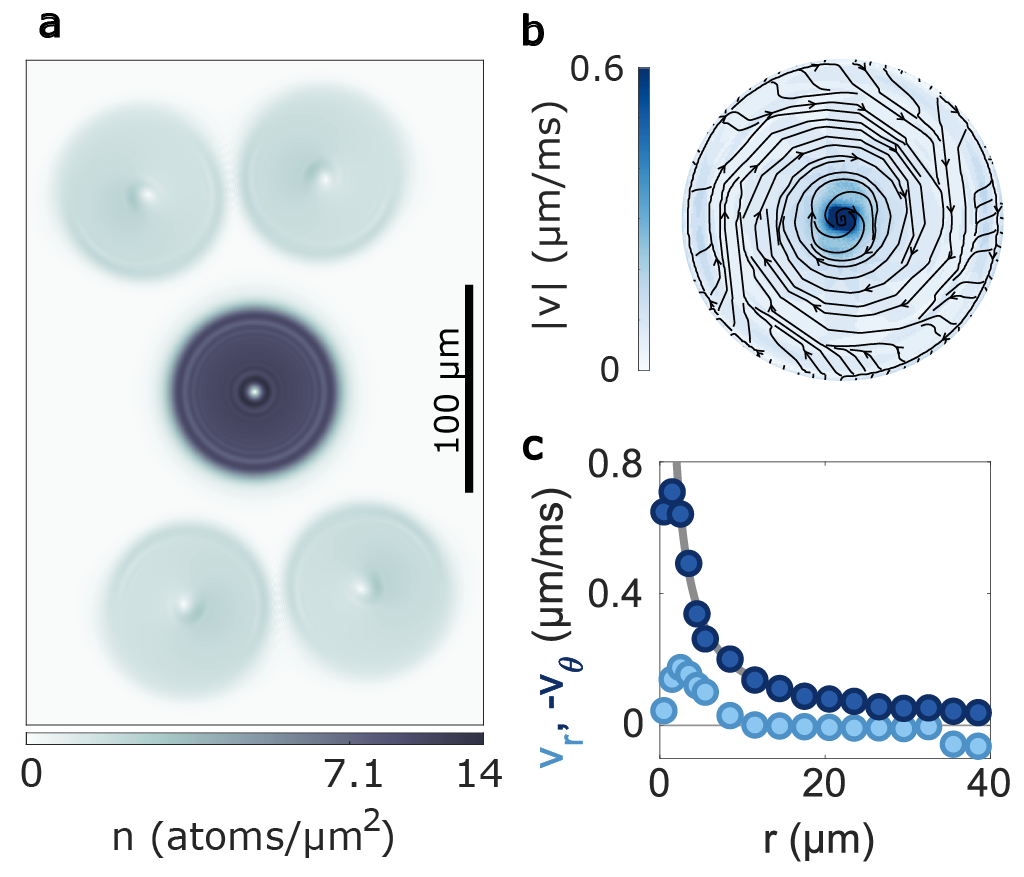} 
     \caption{\textbf{Result of a Velocity Field Reconstruction Simulated with the Gross-Pitaevskii Equation.} 
     \textbf{a)} Simulated density distribution. The density is displayed using nonlinear color scaling. 
     \textbf{b)} Reconstructed in-plane velocity magnitude, with black streamlines indicating the direction of the velocity field. 
     \textbf{c)} Azimuthally averaged radial velocity $v_r$ (light blue) and azimuthal velocity $-v_\theta$ (dark blue), compared with the theoretical prediction (grey line).}
     \label{fig:Figure_Supp_GPE}
\end{figure}
To investigate the various processes associated with the velocity-sensitive measurement procedure, numerical simulations were performed by solving the quasi-two-dimensional Gross–Pitaevskii equation (GPE) using the split-step Fourier method.

The simulations were performed on a $9600 \times 9600$ spatial grid with a resolution of $(0.0625\,\mu\mathrm{m})^2$ to facilitate the high wavenumbers necessary to capture the momentum transfer process, and a time step of $2\,\mu\mathrm{s}$. A phase winding of $-2\pi$ is imprinted before the ground state of the condensate is obtained through imaginary-time propagation. 

The subsequent real-time evolution follows the experimental sequence. 
Similar physical parameters and conditions were used to those in the experiment.
The resulting density distributions, shown in fig.~\ref{fig:Figure_Supp_GPE}a, were binned to a pixel size of $(0.44\,\mu\mathrm{m})^2$ before applying the same velocity reconstruction procedure as in the experiment. This yields the velocity field shown in \cref{fig:Figure_Supp_GPE}b,c.

The simulation reproduces the trends in $v_r$ near the vortex core and at the edge of the BEC. The non-zero inward velocity present in the experimental data for radii $10\,\mu\text{m}<r<30\,\mu\text{m}$ is attributed to motion induced by small deviations due to fringe patterns on the light fields.

Near the vortex core, the extracted $v_\theta$ agrees better with theoretical expectations than the experimentally obtained one shown in \cref{fig:Fig3}. This difference is attributed to the higher spatial resolution of the simulation and the absence of fluctuations in the vortex position.
\\\quad\\
\subsection*{De-Noising Single-Shot Velocity Fields}
\begin{figure*}[b]
    \centering
    \includegraphics{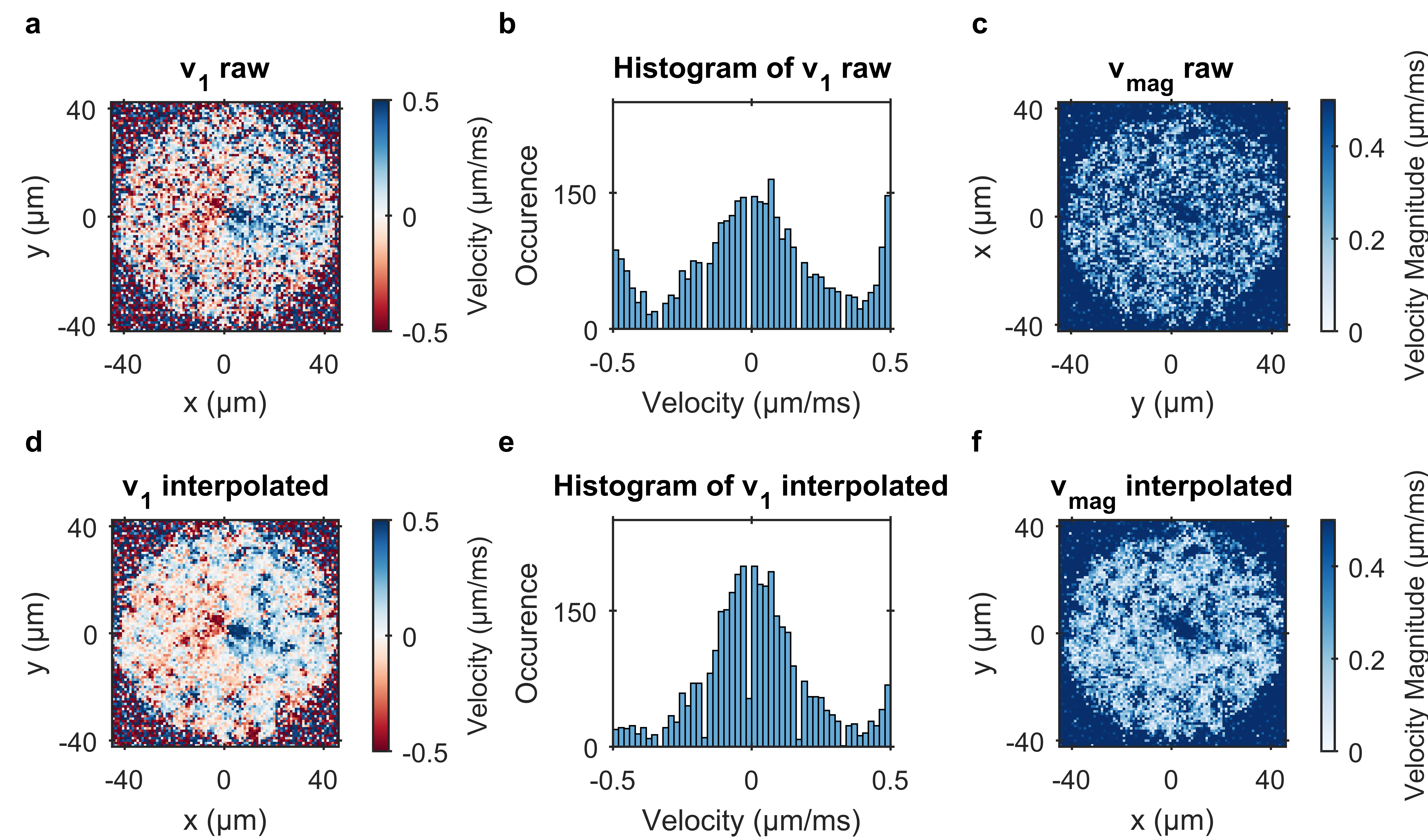} 
    \caption{\textbf{Interpolation of Spurious Pixels in Single-Shot Velocity Fields.} 
    \textbf{a)} As velocity field extraction depends on ratios of atomic densities, the method is susceptible to noise in the camera images. Noisy pixels lead to isolated spurious data points with high velocity values in the extracted fields $v_1$ and $v_2$. 
    \textbf{b)} In the histogram of extracted velocities $v_1$ inside the cloud, these spurious points pile up at the edges of the distribution. 
    \textbf{c)} Since the spurious data points in $v_1$ and $v_2$ are not necessarily located at the same positions, they can significantly distort the total velocity magnitude $v_{\mathrm{mag}}$. 
    \textbf{d)} Spurious points are identified in each velocity component using a local median absolute deviation outlier rejection, and filled in by interpolation using a biharmonic inpainting. 
    \textbf{e)} The histogram of $v_1$ after interpolation shows fewer pixels with high velocities. 
    \textbf{f)} Consequently, $v_{\mathrm{mag}}$ appears smoother.}
     \label{fig:FigSuppInfill}
\end{figure*}

Our velocity reconstruction relies on density ratios between five individual pixels. These ratios are sensitive to noise. This becomes particularly problematic when extracting velocity fields from single experimental shots, where the reduced signal-to-noise-ratio can produce isolated, spurious data points with velocities unphysically high when compared to their surroundings (\cref{fig:FigSuppInfill}, a to c).
To mitigate this effect, the density data is first binned in $2\times2$ camera pixels before reconstructing the velocity fields. 
The unphysical pixels are identified as outliers with respect to the local median and filled in using biharmonic interpolation (\cref{fig:FigSuppInfill}, d to f).

\end{document}